\documentclass[conference]{IEEEtran}
\IEEEoverridecommandlockouts
\usepackage{cite}
\usepackage{amsmath,amssymb,amsfonts}
\usepackage{algorithmic}
\usepackage{graphicx}
\usepackage{textcomp}
\usepackage{xcolor}
\usepackage{url}
\usepackage{booktabs}
\usepackage{caption} 
\usepackage{tabularx}

\usepackage{cleveref}
\usepackage{setspace}
\usepackage{algorithmic,float}
\usepackage{varwidth}
\usepackage[algoruled,nofillcomment,vlined,linesnumbered]{algorithm2e}
\usepackage[inline]{enumitem}
\usepackage{multirow}

\usepackage{color, colortbl}
\definecolor{Gray}{gray}{0.9}
\usepackage[draft,mode=multiuser]{fixme}

\FXRegisterAuthor{dc}{andc}{DC}
\FXRegisterAuthor{am}{anam}{AM}
\FXRegisterAuthor{ar}{anar}{AR}
\FXRegisterAuthor{lv}{anlv}{LV}
\fxsetup{layout={marginclue,footnote}}
\fxusetheme{colorsig}

\newcommand{\toolname}{3PDroid}

\def\BibTeX{{\rm B\kern-.05em{\sc i\kern-.025em b}\kern-.08em
    T\kern-.1667em\lower.7ex\hbox{E}\kern-.125emX}}

\title{On the (Un)Reliability of Privacy Policies in Android Apps}

\begin{document}

\author{\IEEEauthorblockN{Luca Verderame$^{1}$, Davide Caputo$^{1}$, Andrea Romdhana$^{1,2}$, Alessio Merlo$^{1}$}
\IEEEauthorblockA{\textit{$^1$DIBRIS}, \textit{University of Genoa}, Genoa, Italy \\
Email: \{luca.verderame, davide.caputo, andrea.romdhana, alessio\}@dibris.unige.it}

\IEEEauthorblockA{\textit{$^2$Security \& Trust Unit}, \textit{FBK-ICT}, Trento, Italy }
%Email: aromdhana@fbk.eu}
}
%\and
%\and
%\and
%\and
%\and

\maketitle

\begin{abstract}
The access to privacy-sensitive information on Android is a growing concern in the mobile community. Albeit Google Play recently introduced some \textsl{privacy guidelines}, it is still an open problem to soundly verify whether apps actually comply with such rules. To this aim, in this paper, we discuss a novel methodology based on a fruitful combination of static analysis, dynamic analysis, and machine learning techniques, which allows assessing such compliance. More in detail, our methodology checks whether each app i) contains a privacy policy that complies with the Google Play privacy guidelines, and ii) accesses privacy-sensitive information only upon the acceptance of the policy by the user. Furthermore, the methodology also allows checking the compliance of third-party libraries embedded in the apps w.r.t. the same privacy guidelines. \\
We implemented our methodology in a tool, \toolname, and we carried out an assessment on a set of recent and most-downloaded Android apps in the Google Play Store. Experimental results suggest that more than 95\% of apps access user's privacy-sensitive information, but just a negligible subset of them ($\approx 1\%$) fully complies with the Google Play privacy guidelines.
\end{abstract}


\section{Introduction}
\label{sec:intro}
According to Statista\footnote{https://www.statista.com/statistics/266210/number-of-available-applications-in-the-google-play-store/}, the number of available Android applications (hereafter, apps) was lowering in 2019, for the first time. This fact suggests that the app market competition is becoming more fierce, where a lot of apps drop from the Google Play Store due to obsolescence or lack of interest by the users' community. In order to stay on top, apps need to keep monitoring the users' preferences and demands, by continuously harvesting and gathering both user's and device information during their execution. 
Unfortunately, most of such information are privacy-related (e.g., the user's location through GPS, the contact list, and the IMEI of the device), and raised privacy concerns from both corporate and personal users (see, e.g., \cite{privacy_social,wang2019implications}). This is due to the recent discovery of severe data breaches involving mobile apps, like the one discovered in the Peekaboo Moments Android app which exposed more than 100 GB of images and videos of babies\footnote{https://www.bankinfosecurity.com/babys-first-breach-app-exposes-baby-photos-videos-a-13603}.

To deal with users' privacy demand, mobile apps should only access the minimum amount of \textsl{personal and sensitive information} (hereafter, PSI) which could be sufficient to provide the service they are offering. Furthermore, they should clearly state which PSI is accessed. 

To try mitigating the privacy problem on the Android platform, the Google Play Store released a detailed document containing a set of \textsl{privacy guidelines}
%\amnote{Ho chiamato la privacy policy di Android come Privacy Guidelines of the Google Play Store. Please se ha un nome milgiore cambiatelo e fatelo throughout the introduction.} 
for Android apps \cite{PrivacyPolicyGooglePlayStore}. This document contains the technical and legal requirements concerning \textsl{``the collection, use, and sharing of the data, and limiting the use of the data to the purposes disclosed, and the consent provided by the user}''. As a consequence, all apps on the Google Play Store that need to access PSI must have a \textsl{privacy policy} to notify the user about how they collect, use, share, and process such information. It is also mandatory that this policy is hosted inside the app, and is prompted to (and accepted by) the user before accessing any PSI.

%Currently, there are no automatic methodologies capable of identifying whether an app complies with the new Google Play Privacy guidelines.
%Recent work has begun studying whether or not mobile app behavior matches statements in privacy policies e.g. \cite{Yu2016,Story2018,Zimmeck2017}.
%Unfortunately, those controls are not enough to determine the privacy compliance of the apps. Indeed, most of the proposals rely only on static analysis techniques and focus on the privacy policy page extracted from the Google Play website rather then from the app itself. In addition to that they do not provide any solution \emph{i)} to determine if the privacy policies page in the app adeheres with the Google Play Guidelines 
%The new regulations, indeed, describe the way how the developers should include the privacy policy page in the app, and one of the first points of these regulation says: "the privacy policy page must be inside app, and not only in a web page or in the app page on Google Play".
%Furthermore, the extensive usage of third-party libraries and frameworks \cite{privacy_thirdparty} boosts the complexity in assessing the type of information that is collected and processed by the app.

\paragraph*{\textbf{Research challenge}} Albeit the definition of a set of privacy guidelines is an important step towards providing privacy guarantees to Android users, a sound and complete methodology to assess whether each app actually complies with them is still missing.  Furthermore, it is likewise unclear how many apps actually comply with the Google Play privacy guidelines.

%\amnote{Da precisare meglio, come punti. In questa direzione noi facciamo...}

\subsection{State of the art} 
\label{sec:sota}
In 2018, Google carried out an extensive analysis in the Google Play Store, thereby removing all apps that did not have at least an external link to a valid privacy policy page on their Google Play Store page. To prevent Google from deleting their apps, several developers added such link, bringing the percentage of apps that satisfies this requirement from $41.7\%$ in 2017 to $51.8\%$ in 2018 \cite{Story2018}.
However, no further checks on the content of the privacy policy page have been carried out, nor on the compliance between the privacy policy and the actual behaviors of the app.%\\
%To overcome the study of the correlation between the privacy policy and behavior of the mobile app is one of the most studied topics by the scientific community.

To deal with previous issues, Zimmeck et al. \cite{Zimmeck2017} carried out an extensive analysis of $17.991$ free Android apps on the Google Play Store by leveraging static analysis and machine learning techniques. The authors found that $71\%$ of apps that lack a privacy policy should have one, as they access PSI. 
Also, for $9,050$ apps that have a privacy policy page, the authors found a lot of  \textsl{potential} inconsistencies between the content of the policy page and the app behavior. Unfortunately, since the authors relied on static analysis only, the actual number of true positives cannot be verified.  In \cite{Zimmeck2019}, the authors proposed a tool, named Mobile App Privacy System (MAPS), which is able to carry out a more extensive analysis of Android apps. MAPS is based on a pipeline for retrieving and analyzing large app populations based on code analysis and machine learning techniques. 
MAPS analyzed $1,035,853$ apps taken from the Google Play Store, and found that only $50.5\%$ of them actually have a privacy policy page.
In \cite{Ando2019PolicyLint}, the authors propose PolicyLint, a privacy policy analysis tool able to identify contradictions inside the privacy policy page. The authors analyzed $11,430$ apps and found that $14.2\%$ of the privacy policies contain contradictions that may suggest the presence of misleading statements.
\paragraph*{\textbf{Open challenges}} 
We argue that previous proposals suffer from some limitations. First, they rely on static analysis techniques only, thereby making hard to identify actual true positives. Moreover, they focus on the privacy policy page published on the Google Play Store only, which may differ from the one contained in the app and prompted to the user at runtime. Furthermore, they are not able to \emph{i)} identify the privacy policy page inside the app, \emph{ii)} verify whether such page complies with the privacy guidelines of the Google Play Store, \emph{iii)} detect whether the app begins to access PSI \textsl{before} the user explicitly accepts the privacy policy.\\
Previous challenges require to assess the behavior of the app \textsl{dynamically} and are getting worse by the fact that recent
apps made extensive usage of third-party libraries and
frameworks \cite{privacy_thirdparty}, which boost the complexity in assessing the type of data that is collected and processed by the app.
%Indeed, most of the proposals rely only on static analysis techniques and focus on the privacy policy page extracted from the Google Play website rather than from the app itself. 
%In addition to that they do not provide any solution 
%\lvnote{questa frase forse è un po scollegata da sopra ma va secondo me puntualizzata somewhere}

\paragraph*{\textbf{\toolname}} In this paper, we propose a novel methodology to automatically verify the compliance of an Android app with the privacy guidelines of the Google Play Store and we implemented it in a tool, named \toolname\footnote{The tool and the results are available at  \underline{\url{https://csec.it/3pdroid}}.} This tool  combines static and dynamic analysis approaches with machine learning techniques to monitor the runtime behavior of an Android app in order to verify whether:
\begin{enumerate}
   \item the app contains a privacy policy page;
   %\item the content of the privacy policy page complies with the privacy guidelines of the Google Play Store;
   \item the privacy policy page fully complies with the Google Play privacy guidelines; 
   \item the app accesses PSI \textsl{only after} the user has accepted the privacy policy;
  \item the access to PSI - carried out by both the native app and its included third-party libraries/frameworks - complies with the app privacy policy.
\end{enumerate}
\vspace{0.1in}

\paragraph*{Structure of the paper} The rest of the paper is organized as follows: Section \ref{sec:background} introduces some basics on Android and the Google Play privacy guidelines, while Section \ref{sec:methodology} presents a methodology to automatically assess the compliance of Android apps with the privacy guidelines. Section \ref{sec:implementation} discusses the implementation of the proposed methodology in \toolname{}, as well as an experimental setup aimed at systematically analyzing sets of Android apps. Section \ref{sec:experimental_results} shows and discusses the experimental results, while Section \ref{sec:conclusion} concludes the paper and points out some extension of the work.
%\lvnote{organizzazione del paper}
%\lvnote{TODO qui: 
%1) capire se espandere un pò di più i singoli punti del contributi;
%2) capire se vogliamo sbilanciarci in termini di PoC o use-case con una frase
%2b) @ale capire se vogliamo già fare commitment suggerendo che faremo degli esperimenti
%3) mettere l'organizzazione dell'app
%}

\section{Background}
\label{sec:background}
% Nel mondo mobile --> tipo di dati sensibili (cite qualche lavoro) --> android e permessi per accedere a questo tipo di dati e per accedere a questi dati devo utilizzare specifiche api --> di conseguqenza è possibile fare un mapping tra le i PII (personally identifiable information) - permessi richiesti e API.
% Sfortunatamente però oltre ai classici PII la privacy dell'utente è messa a rischio dalla possibilità di profilare utenti etc (video etc)
% Questo viene fatto con librerie di terze-parti analytics
% Sfortunatamente questo tipo di dato non essendo un PII standard di Android non è soggetto a nessun meccanismo di defautl di restrizione dato dal SO
% una conseguenza una metodologia di analisi dovrebbe considerare entrambi questi aspetti

Modern mobile devices gather a plethora of PSI, which could refer to different categories, like financial and payment data, authentication information, phonebook, contacts, SMS, call-related data, microphone, and camera sensor data, just to cite a few. % \lvnote{probabilmente visto che personal and sensitive information è importante conviene introdurre il termine in intro}
Android apps extensively collect, store, and process PSI in order to improve their quality and reliability, as extensively discussed in \cite{Grace2012Unsafe,Huber2013AppInspect,Ren2016ReCon}. %\lvnote{TODO valutare se togliere reference}
From a technical standpoint, apps can retrieve PSI either by i) relying on the Android platform API or ii) embedding third-party libraries.

\paragraph*{\textbf{Android API and Permissions}} 
The Android OS allows apps to get access to PSI through a set of well-defined API. At the same time, Android limits the access to PSI through a security mechanism based on the idea of \textsl{permission}. In a nutshell, each PSI-related API is associated with a set of privacy-sensitive permissions. The invocation of an API is then restricted to the sole apps having the required set of permissions. Android apps must require permissions by declaring them in their \textit{AndroidManifest.xml} file in a specific tag named \textit{uses-permission}\cite{AndroidManifest}. 

According to the type of permission, the OS might grant it automatically or might prompt the user to approve the request. For example, if an app aims to read data from the contact list, it must declare the ``\texttt{android.permission.READ\-\_CONTACTS}" permission within the \textit{AndroidManifest.xml} file in order to be allowed to invoke the corresponding API (e.g., \texttt{getPhoneNumbers}). Then, the first time the app tries to access the API, the user is prompted with a permission grant request. If the user accepts, the app can invoke the API, thereafter. 

For the aim of this work, we mapped the standard permission set provided by the Android OS with the corresponding API methods and the PSI. An excerpt of such mapping, inspired by the works of \cite{Reyes2018Coppa} and \cite{Ren2016ReCon}, is provided in Table \ref{tab:example_of_psi}. %\lvnote{possiamo lasciare solo 2 reference rilevanti?}
\begin{table}[h]
\footnotesize
\centering
\begin{tabularx}{250pt}{lll}
\toprule

\textbf{Permission} &\textbf{PSI} & \textbf{API method}  \\
    \midrule
    % \texttt{READ\_CALENDAR} & Calendar & \texttt{Uri.parse(``content://com.android.calendar")}\\
    % \texttt{READ\_CONTACTS} & Contacts & \texttt{Uri.parse(``content://contact")}\\
    \texttt{ACCESS\_NETWORK\_STATE} & Device MAC & \texttt{getMacAddress()}\\
    \texttt{ACCESS\_FINE\_LOCATION} & GPS Location & \texttt{getLocation()} \\
    \texttt{READ\_PHONE\_STATE} & IMEI & \texttt{getDeviceId()}\\
    \texttt{READ\_PHONE\_STATE} & Phone Number & \texttt{getLine1Number()}\\
    \texttt{ACCESS\_WIFI\_STATE} & Router MAC & \texttt{getMacAddress()}\\
    \texttt{ACCESS\_WIFI\_STATE} & Router SSID &  \texttt{getBSSID()} \\
    % \texttt{READ\_SMS} & SMS Message & \texttt{Uri.parse(``content://sms/inbox")}\\

\bottomrule
\end{tabularx}
\caption{Mapping of Android permissions, PSI and API methods.}
%\dcnote{da rivedere questa tabella per i metodi}
\label{tab:example_of_psi}
\end{table}

\paragraph*{\textbf{Third-party Libraries}}
The previous mapping provides only a partial coverage of the full PSI. In fact, there exists a wide set of PSI that is beyond the control of the OS and, therefore, it is outside the enforcement mechanisms of the Android permissions. 
Examples of this PSI include all the usage statistics like the commercial or usage preferences of the user, collected for profiling aim, or to build crash reporting activities. To collect this PSI, developers extensively include in their apps third-party libraries for analytics and advertising that can be triggered using a set of the API methods\cite{HE2019259}.

For instance, analytics libraries can retrieve and store several pieces of PSI, including the country as well as the type and the model of the mobile device.  Most of these libraries can also track the user's activities. % or the performance of advertisement networks.
Furthermore, developers may use \textsl{Ad Network} libraries to deliver application ads in order to increase revenues. To this aim, the Ad Network library may access PSI to show personalized ads customized to the taste of the users.
Notables examples of analytics and Ad Network libraries include AdMob\footnote{https://admob.google.com/intl/it/home/}, Google Analytics\footnote{ttps://firebase.google.com/docs/analytics}, InMobi\footnote{https://www.inmobi.com/}, and Facebook Analytics\footnote{ttps://analytics.facebook.com/}.

%From a technical standpoint, third-party libraries offer a set of API that can be triggered by the application to exploit a specific functionality.
In this work, we identified a set of the most used analytics and Ad Network libraries according to \cite{Book2013CollusionAd}, and we mapped the API methods that involve the collection, process, and use of PSI.
An excerpt of such mapping, presenting a subset of InMobi API methods, is presented in Table \ref{tab:example_of_libraries}.
%\lvnote{TODO rimpicciolire tabella usando lo stesso formato della prima}

\begin{table}[h]
\footnotesize
\centering
\begin{tabularx}{0.40\textwidth}{lX}
\toprule

\textbf{API method} & \textbf{PSI}  \\
    \midrule
    \texttt{setKeywords} & Keywords \\
    \texttt{setSearchString} & Keywords  \\
    \texttt{setGender} & Gender  \\
    \texttt{setCurrentLocation} & Location  \\
    \texttt{setAge} & Age  \\
    \texttt{setRequestParams} &  Multiple Factors  \\
    \texttt{setPostalCode} & Postal Code  \\
     \texttt{setLocationInquiryAllowed}  & Enable Location \\
    \texttt{setIncome} & Income  \\
    \texttt{setInterests} & Interests   \\
    \texttt{setAreaCode} & Area Code  \\
    \texttt{setEducation} & Education   \\
    \texttt{setEthinicity} & Ethnicity   \\

\bottomrule
\end{tabularx}
\caption{Examples of API methods extracted from the InMobi library and their mapping with PSI.}
\label{tab:example_of_libraries}
\end{table}

% STATS --> https://arxiv.org/abs/1307.6082
% TESI MATTIA

\subsection{The Google Play Privacy Guidelines}\label{sec:google_play_store_regulations} 

The growing concerns on PSI pushed Google Play to release a ``Privacy, Security and Deception" guideline document \cite{PrivacyPolicyGooglePlayStore} to grant transparency on the access and usage of PSI by Android apps.
Such guidelines force each developer to restrict the collection and the use of sensitive data only for aims directly related to the supply and improvement of the functionality of the app. Furthermore, the developer must handle all of these pieces of data safely and transmit them using modern encryption mechanisms (i.e., via HTTPS).

Still, in case PSI are gathered at runtime, the app must provide an in-app disclosure regarding data collection and usage, i.e., a \textsl{privacy policy page}. Such page must meet a set of \textsl{technical} (\emph{TR}) and \textsl{content} (\emph{CR}) requirements, as detailed in Table \ref{tab:google_play_guidlines}, in order to be compliant with the guidelines. 
In a nutshell, the application must include the policy within the app, and prompt it to the user without requiring her to open a menu or the settings.  Moreover, the app must require the explicit consent of the user, e.g., by avoiding timeout or automatic acceptance actions.
Finally, the privacy policy page must precisely describe both the set of collected PSI and for which aim. At the same time, the app must not collect PSI before obtaining the user's consent.

\begin{table}[h]
\footnotesize
\centering
\begin{tabularx}{250pt}{lX}
\toprule

\textbf{Id} & \textbf{Description} \\
    \toprule
    \texttt{CR1} & The privacy policy page must clearly state the set of PSI it collects, as well as how the same PSI will be used.\\
    \midrule
    % CR2 & The privacy policy page must explain how the data will be used \\
    % \midrule
    \texttt{CR2} & The privacy policy page must be prompted to the user in a proper document, which is different from the terms of service or other documents that do not deal with personal or sensitive information. \\
    \midrule 
    % CR3 & The privacy policy page cannot be included exclusively in the privacy policy or in the Terms of Service \\
    % \midrule
    % CR4 & The privacy policy page cannot be included in other information not related to the collection of personal or sensitive data \\
    % \midrule
    % CR5 & The privacy policy page must present the consent dialog clearly and unequivocally \\
    % \midrule
    \texttt{TR1} & The privacy policy page must be stored within the app, i.e., it is not sufficient to store it on a website or on the app page in the Google Play Store. \\
    \midrule
    \texttt{TR2} & The privacy policy page must be shown during the execution of the app once and automatically prompted, i.e., the user must not search for it in a menu or a settings page.\\
    \midrule
    \texttt{TR3} & The user must explicitly accept the privacy policy, e.g., it must click on an acceptance widget.\\ 
    \midrule
    \texttt{TR4} & The app must not collect PSI before the user accepts the privacy policy. \\
    \midrule
    \texttt{TR5} & The app must not consider the privacy policy as accepted if the user leaves the privacy policy page by pressing the home or back button. \\ 
    \midrule
    \texttt{TR6} & Once prompted, the privacy policy page must not expire before the user could accept it.\\
\bottomrule
\end{tabularx}
\caption{Technical (TR) and content (CR) requirements of the Google Play privacy guidelines.}
\label{tab:google_play_guidlines}
\end{table}

%\lvnote{@davide aggiungi la tabella copiando da qui https://play.google.com/intl/en_us/about/privacy-security-deception/user-data/#!?zippy_activeEl=personal-sensitive#personal-sensitive}

% \subsection{Natural Language Processing}
% \amnote{Qui tutto quello che serve per la parte di ML che usiamo in CiaoDroid.}

% \amnote{NOTA BENE: l'output di questa sezione e' tutto cio' che serve per un lettore per poi capire la parte di methdology. }
\section{Assessing the privacy guidelines compliance}
\label{sec:methodology}
\begin{figure}
    \centering
    \includegraphics[scale=0.37]{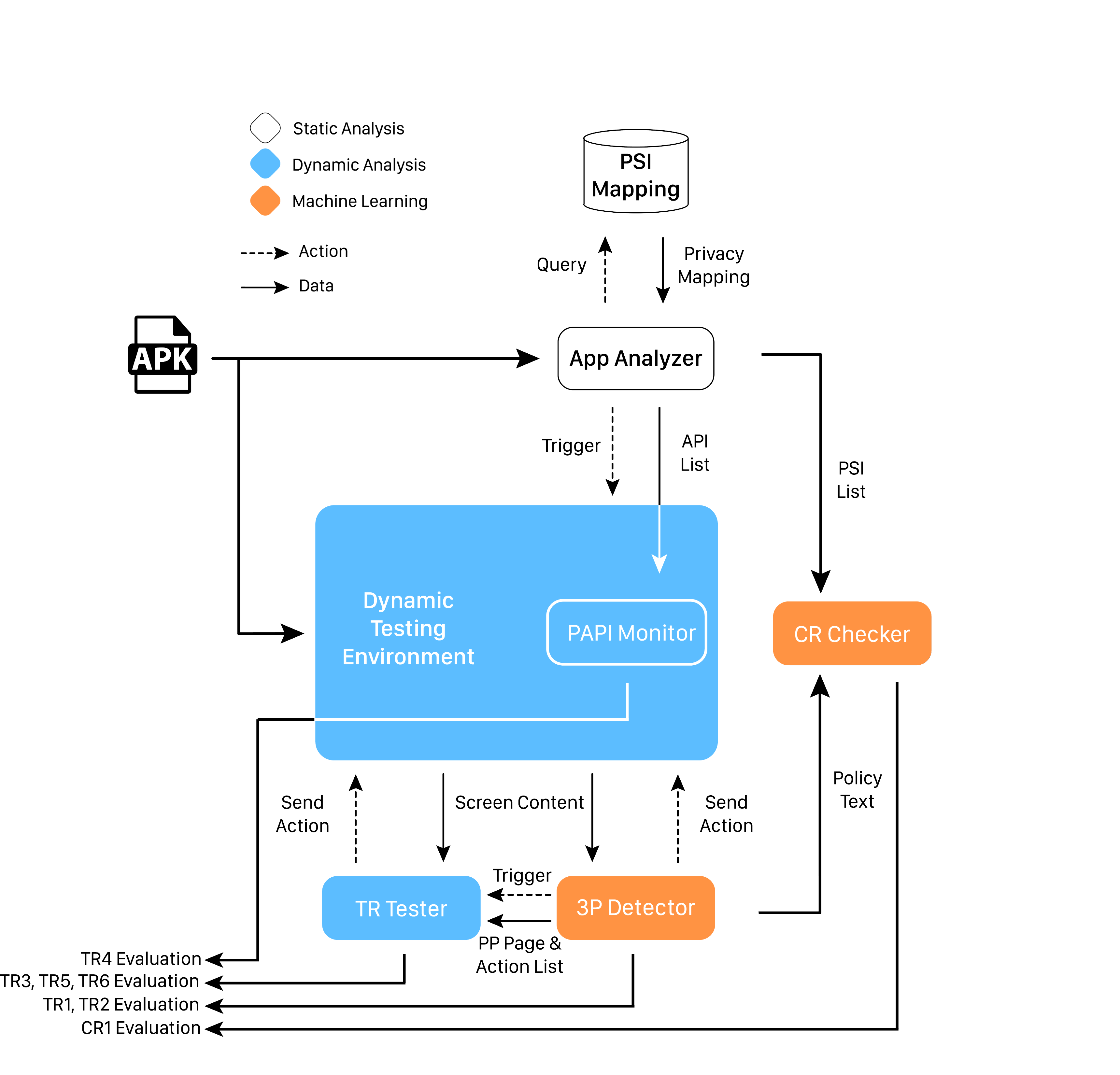}
    \caption{Analysis Workflow.}
    \label{fig:figure_workflow}
\end{figure}

We discuss here a novel methodology based on a combination of static analysis, dynamic analysis, and machine learning techniques to address the open research challenges described at the end of Section \ref{sec:intro}. This methodology can be applied to any Android application package without requiring the source code nor any additional information, and it is composed of $7$ different modules that cooperate according to the workflow sketched in Figure \ref{fig:figure_workflow}. The rest of this section details the modules and their interactions. 

\subsection{PSI Mapping}

The \textit{PSI Mapping} contains the API methods belonging to both the Android OS and the third-party libraries that are relevant for the analysis, i.e., that collect, use, or process PSI.
In detail, the \textit{PSI Mapping} contains the mapping of all the privacy-related Android permissions (see Table \ref{tab:example_of_psi}) and the third-party libraries (see Table \ref{tab:example_of_libraries}) with their corresponding PSI-related API methods.
%, and \emph{ii)} third-party libraries and their corresponding PSI-related API methods . %, and it is organized per Android API level and library version,.

\subsection{App Analyzer}
% \dcnote{Toglierei la categoria delle app? Anche perchè come potremmo giustificarne l'uso?}
%The module \textit{App Information Extractor} is the first of the entire workflow. 
%The goal of this module is to extract information from the app such as the list of permission requested, and analytics and ads libraries used. %and the app's category.

%If the app does not require any type of privacy-sensitive permission and does not use any third-party libraries, it is not forced to have a privacy policy page, so the app is marked as compliant with the Google Play regulations.
%All this information is passed to the next module: \textit{APIs Monitor}.

The \textit{App Analyzer} decompiles an Android application package (APK) and extracts \emph{i)} the list of permissions requested by the app and \emph{ii)} the list of third-party libraries included in the app.
For each finding, the module queries the \textit{PSI Mapping} to determine if the permission or the library is privacy-sensitive. If this is the case, it requests the PSI and the corresponding list of API methods that are involved.
Then, the \textit{App Analyzer} dispatches the list of PSI to the \textit{Content Requirement Checker (CR Checker)} and the list of involved API methods to the \textit{Privacy API Monitor (PAPI Monitor)}. %, for the remaining of the analysis. 
Finally, the module triggers the \textit{Dynamic Testing Environment (DTE)} for the dynamic analysis phase.
Instead, if the app does not contain any privacy-related permission or library, it is marked as compliant and no further analyzed.

\subsection{Dynamic Testing Environment}

%In order to dynamically evaluate the privacy compliance of the app, the methodology executes a set of tests using dynamic analysis techniques.
%The \textit{Dynamic Testing Environment (DTE)} is the module in charge of the installation and the execution of the Android app inside a device emulator. Firstly, the \textit{DTE} initializes an Android emulated device, then it installs and starts the app. 
%and iii) triggers the other dynamic modules (namely, the API monitor, the PP Tester, and the PP Identifier) to start the analysis.

The \textit{Dynamic Testing Environment (DTE)} is the module in charge of the installation and the execution of the app inside a device emulator. First, the DTE initializes an Android emulated device and installs the APK.
Then, the module executes the app, and notifies both the PAPI Monitor and the \textit{Privacy Policy Page Detector (3P Detector)} to start the analysis.

\subsection{Privacy Policy Page Detector (3P Detector)}

The goal of the \textit{Privacy Policy Page Detector (3P Detector)} module is to identify the privacy policy page screen inside the Android app.
First, the 3P Detector connects to the DTE and searches for the privacy policy page.

Following the algorithm depicted in Algorithm \ref{alg:privacypolicypagedetector}, the module retrieves the XML content of the app screen which is currently displayed to the user (row 5). Then, the module aims to determine whether the textual content of the extracted page includes a privacy policy (row 6-8).
To do so, the 3P Detector enforces a text classification algorithm based on machine learning techniques.
The classification procedure requires two phases, namely \emph{preprocessing} and \emph{classification}. 

\paragraph*{\textbf{Preprocessing}}
At first, the preprocessing extracts the text content (row 6) from the XML of the page. Then, the extracted text is polished to normalize whitespace and punctuation, remove non-ASCII characters, and lowercase all the content (row 7).

\paragraph*{\textbf{Classification}}
The preprocessed text is then evaluated to determine if it contains a privacy policy page. For such a task, 3P Detector relies on a multilayer perceptron classifier (row 8).
If the classifier marks the text as a privacy policy, the 3P Detector sets the flag \texttt{detected} and exits the research loop.
Otherwise, the module sends an input to the app (e.g., a swipe or a press command) in order to evaluate a new screen at the next iteration (row 12-13). Also, 3P Detector stores a list of the input actions (row 14) used to reach the privacy policy page. This information will be sent to the \textit{Technical Requirement Tester} to evaluate the list of \texttt{TRs} (see Table \ref{tab:google_play_guidlines}).
Finally, if the module successfully detects the policy page, it returns the text of the policy, the XML file, and the list of actions required to reach the page (row 18). 
On the contrary, if the module cannot find the policy page within a maximum limit of input actions (i.e., \textsl{MA}), the app is marked as not compliant, and the analysis terminates. 
Indeed, since apps must satisfy \texttt{TR2}, the module expects to reach the policy page within such a threshold.
%\lvnote{TODO aggiungere in pseudocode le due funzioni per il pre-processing e la classification in modo da aggiornare il testo}

\begin{algorithm}

            \footnotesize
            %\setstretch{0.5}
    
            \caption{3P Detector}
            \SetAlgoLined
            \label{alg:privacypolicypagedetector}
            \SetKwData{emulator}{emulator}
            \SetKwData{pageContent}{pageContent}
            \SetKwData{nextAction}{nextAction}
            \SetKwData{contentPrivacyPolicyPage}{contentPrivacyPolicyPage}
            \SetKwData{screenshotPrivacyPolicyPage}{screenshotPrivacyPolicyPage}
            \SetKwData{resultExperiment}{resultExperiment}
            \SetKwData{result}{result}
            \SetKwData{listAPIsInvoked}{listAPIsInvoked}
            \SetKwData{privacyPolicyPage}{privacyPolicyPage}
            \SetKwData{pageHomeButton}{pageHomeButton}
            \SetKwData{pageBackButton}{pageBackButton}
            \SetKwData{timeout}{timeout}
            \SetKwData{MaxActions}{MA}
            \SetKwData{notCompliant}{notCompliant}

            \SetKwData{listActions}{listActions}
            \SetKwData{detected}{detected}
            \SetKwData{xmlContent}{xmlContent}
            \SetKwData{textContent}{textContent}
            \SetKwData{preprText}{preprText}
            \SetKwData{MLPClassifier}{MLPClassifier}
            \SetKwData{PPPXml}{PPPXml}

            %% FUNCTION    
            \SetKwFunction{startAPK}{startAPK}
            \SetKwFunction{getPageContent}{getPageContent}
            \SetKwFunction{isPolicyPage}{isPolicyPage}
            \SetKwFunction{getNextAction}{getNextAction}
            \SetKwFunction{performAction}{performAction}
            \SetKwFunction{getContentPrivacyPolicyPage}{getContentPrivacyPolicyPage}
            \SetKwFunction{takeScreenshot}{takeScreenshot}
            \SetKwFunction{containsExplAccept}{containsExplAccept}
            \SetKwFunction{getAPIsInvoked}{getAPIsInvoked}
            \SetKwFunction{setListAPIsInvoked}{setListAPIsInvoked}
            \SetKwFunction{makeList}{makeList}
            \SetKwFunction{add}{add}
            
            \SetKwFunction{len}{len}
            \SetKwFunction{setResult}{setResult}
            \SetKwFunction{setContentPrivacyPolicyPage}{setContentPrivacyPolicyPage}
            \SetKwFunction{wait}{wait}
            \SetKwFunction{pressHomeButton}{pressHomeButton}
            \SetKwFunction{pressBackButton}{pressBackButton}
            \SetKwFunction{startAppWithPrivacyPolicyPage}{startAppWithPrivacyPolicyPage}
            \SetKwFunction{extractFromXML}{extractFromXML}
            \SetKwFunction{dataPreprocessing}{dataPreprocessing}
            
            \SetKwInOut{Input}{Input}\SetKwInOut{Output}{Output}
            \BlankLine
            \Input{Emulator, \MaxActions}
            \Output{textContent, PPPXml, listActions}
            %\Output{ResultExperiment}
            \Output{TR1-TR2 Evaluation}
            \BlankLine
            \listActions $\gets$ \makeList()\;
            \detected $\gets$ \KwSty{False}\;
            \emulator.\startAPK()\;
            \While{\KwSty{not} \detected \KwSty{and} \listActions.\len() $<$ \MaxActions}
            {
                \xmlContent $\gets$ \emulator.\getPageContent()\;
                \textContent $\gets$ \extractFromXML(\xmlContent)\;
                \preprText $\gets$ \dataPreprocessing(\textContent)\;
                \detected $\gets$ \MLPClassifier.\isPolicyPage(\preprText)\;
                \If{\detected}
                {
                    \KwSty{break};
                }
                \nextAction $\gets$ \getNextAction(\pageContent)\;
                \emulator.\performAction(\nextAction)\;
                \listActions.\add(\nextAction)\;
            }
            \BlankLine
            \If{\detected}
            {
                \BlankLine
                \PPPXml $\gets$ \xmlContent\;
                \Return \textContent, \PPPXml, \listActions;
            }
            \BlankLine
            %\Return \KwSty{\texttt{False}};
            \Return \KwSty{\texttt{Fail\_TR1\_TR2}};
    \end{algorithm}

\subsection{Privacy API Monitor (PAPI Monitor)}

%This module aims to identify that the app does not invoke any privacy relevant API before that the user explicitly accepts the privacy policy page.
%To do that, \textit{Privacy API Monitor (PAPI Monitor)} firstly retrieves the list of privacy relevant APIs from \textit{App Analyzer}, and then inspect that during all the experiment the app does not invoke any API within it (TR4).
%If the app does invoke any of API within this list, it is marked as not compliant.

This module verifies whether the app does invoke any PSI-related API before the user explicitly accepts the privacy policy page.
Once it receives the start notification from the DTE, the \textit{Privacy API Monitor} connects to the testing environment and begins to monitor the execution of any of the PSI-related API method received from the \textit{App Analyzer}.
If the module logs any of these methods, then the app does not fulfill the \texttt{TR4} rule, and thus it is marked as not compliant.

\subsection{Technical Requirement (TR) Tester}

The aim of the \textit{Technical Requirement (TR) Tester} module is to identify whether the privacy policy page satisfies the technical requirements described in Section \ref{sec:google_play_store_regulations}. To do that, the TR Tester implements Algorithm \ref{alg:privacypolicypagealgorithm}.

First, the module evaluates the \texttt{TR3} requirement by analyzing the XML of the page to detect any element allowing the user %to explicit accept the policy 
to accept the policy explicitly (e.g., a button or a checkbox); 
if no match is found, the app is marked as not compliant, and the evaluation terminates (rows 1-3).
Otherwise, the module checks if the page expires or closes automatically within a given threshold (row 4), in order to verify the \texttt{TR6} requirement.
If the policy page is still displayed when the threshold is reached, the module then proceeds with the evaluation of the \texttt{TR5} requirement. To this aim, it triggers the home button and then re-opens the app (rows 8-10): if the app screen is different from the privacy policy page, then the app considers leaving the policy windows as an act of acceptance, and thus it is marked as not compliant (row 14-15). The experiment is repeated with the back button (rows 11-13).

\begin{algorithm}

            \footnotesize
            %\setstretch{0.5}
    
            \caption{TR Tester}
            \SetAlgoLined
            \label{alg:privacypolicypagealgorithm}
            \SetKwData{emulator}{emulator}
            \SetKwData{currentPage}{currentPage}
            \SetKwData{nextAction}{nextAction}
            \SetKwData{contentPrivacyPolicyPage}{contentPrivacyPolicyPage}
            \SetKwData{screenshotPrivacyPolicyPage}{screenshotPrivacyPolicyPage}
            \SetKwData{resultExperiment}{resultExperiment}
            \SetKwData{result}{result}
            \SetKwData{listAPIsInvoked}{listAPIsInvoked}
            \SetKwData{privacyPolicyPage}{privacyPolicyPage}
            \SetKwData{pageHomeButton}{pageHomeButton}
            \SetKwData{pageBackButton}{pageBackButton}
            \SetKwData{timeout}{timeout}
            \SetKwData{PPPXml}{PPPXml}

            \SetKwData{listActions}{listActions}

            %% FUNCTION    
            \SetKwFunction{startAPKWithPage}{startAPKWithPage}
            \SetKwFunction{getCurrentPage}{getCurrentPage}
            \SetKwFunction{isPrivacyPolicyPage}{isPrivacyPolicyPage}
            \SetKwFunction{getNextAction}{getNextAction}
            \SetKwFunction{performAction}{performAction}
            \SetKwFunction{getContentPrivacyPolicyPage}{getContentPrivacyPolicyPage}
            \SetKwFunction{takeScreenshot}{takeScreenshot}
            \SetKwFunction{containsExplAccept}{containsExplAccept}
            \SetKwFunction{getAPIsInvoked}{getAPIsInvoked}
            \SetKwFunction{setListAPIsInvoked}{setListAPIsInvoked}
            \SetKwFunction{setResult}{setResult}
            \SetKwFunction{setContentPrivacyPolicyPage}{setContentPrivacyPolicyPage}
            \SetKwFunction{wait}{wait}
            \SetKwFunction{pressHomeButton}{pressHomeButton}
            \SetKwFunction{pressBackButton}{pressBackButton}
            \SetKwFunction{startAppWithPrivacyPolicyPage}{startAppWithPrivacyPolicyPage}
            \SetKwFunction{getPageContent}{getPageContent}

            \SetKwInOut{Input}{Input}\SetKwInOut{Output}{Output}
            \BlankLine
            \Input{Emulator, PPPXml, listActions}
            %\Output{ResultExperiment}
            \Output{TR3, TR5, TR6 Evaluation}
            \BlankLine
            % $\emulator.\startAPKWithPage(\privacyPolicyPage)$\;
            
            %$\currentPage \gets \emulator.\getCurrentPage()$\;
            \If{\KwSty{not} \PPPXml.\containsExplAccept()}
            {
              %\Return \KwSty{\texttt{False}}; 
              \Return \KwSty{\texttt{Fail\_TR3}};
            }
            \BlankLine
            $\wait(\timeout)$\;
            \BlankLine
            \If{\emulator.\getPageContent() \KwSty{!=} \PPPXml}
            {
                %\Return \KwSty{\texttt{False}};
                \Return \KwSty{\texttt{Fail\_TR6}};
            }
            \BlankLine
            $\emulator.\pressHomeButton()$\;
            $\emulator.\startAPK()$\;
            $\pageHomeButton \gets \emulator.\getPageContent()$\;
            \BlankLine
            $\emulator.\pressBackButton()$\;
            $\emulator.\startAppWithPrivacyPolicyPage()$\;
            $\pageBackButton \gets \emulator.\getPageContent()$\;
            \BlankLine
            \If{\pageBackButton \KwSty{!=} \PPPXml \KwSty{or} \pageHomeButton \KwSty{!=} \PPPXml}
            {
                %\Return \KwSty{\texttt{False}};
                \Return \KwSty{\texttt{Fail\_TR5}};
            }
            \BlankLine
            
            %\Return \KwSty{\texttt{True}};
            \Return \KwSty{\texttt{Pass}}; 
            
        \end{algorithm}

\subsection{Content Requirement (CR) Checker}

The \textit{Content Requirement (CR) Checker} verifies whether the privacy policy page successfully declares all PSI requested by the app, following the \texttt{CR1} requirement.
Given the list of PSI received by the App Analyzer, the module executes several machine learning-based classifiers. %, i.e., one for each PSI.

For this task, CR Checker identifies for each PSI the most meaningful keywords and creates two distinct sets, as suggested in \cite{Zimmeck2017}.
The first one is used to preprocess the policy text and extract all the sentences that contain at least one of the keywords associated with the data type of the PSI under investigation (e.g., for the location PSI, we use terms like `position' or `GPS'). 
On the resulting sentences, CR Checker uses a second set of keywords that refer to the actions available for a given PSI (e.g., for the location PSI, we use terms like `share' or `partner'), to construct unigram and bigram feature vectors \cite{zimmeck2014privee}.

The feature vectors are then used to classify the policy. 
CR Checker uses a set of machine learning models (one for each PSI, see Table \ref{tab:crevaluator_implementation_results}) to determine whether all the pieces of PSI are included in the policy. 
If this is not the case, the module raises a warning regarding the \texttt{CR1} requirement. In this case, the app must pass through a manual inspection phase. %\amnote{Andrea, please verifica se qui e' cambiato qualcosa post implementazione effettiva.}
%\dcnote{Aggiornata}
%\input{sec/Notes_on_implementation.tex}
\section{Implementation and Experimental Setup}
\label{sec:implementation}
We implemented a prototype of our methodology, called \textsl{\toolname{}}, to evaluate both the detection accuracy and the performance on a dataset of Android apps. The rest of this section describes the implementation choices and the corresponding experimental setup.

% \subsection{Notes on Implementation}

%\paragraph*{\textbf{Static Analysis}}
\paragraph*{\textbf{App Analyzer}}
\toolname{}~implements the \textit{App Analyzer} as a Python script based on the Androguard library\footnote{\url{https://androguard.readthedocs.io/}}.
Androguard makes available several APIs allowing to parse the \textit{AndroidManifest.xml} file and retrieve all the privacy-related libraries. 
Furthermore, the \textit{App Analyzer} queries the \textit{PSI Mapping} DB to build the list of API to be monitored during the dynamic analysis phase.

%\paragraph*{\textbf{Database}}
\paragraph*{\textbf{PSI Mapping}}
The \textit{PSI Mapping} is a MongoDB database storing the mappings among PSI, API Methods, Android permissions, and third-party libraries.
This dataset has been built by parsing the Android API reference website\footnote{https://developer.android.com/reference/packages?hl=en} as well as the websites and the documentation provided by third-party developers. The dataset is composed of a set of JSON object documents. 
The current version of the \textit{PSI Mapping} includes the most used libraries according to \cite{AppBrainAnalytics}, like, e.g., AdMob and Google Analytics.

The first column of Table \ref{tab:crevaluator_implementation_results} shows all PSI taken into consideration in this work and inspired by \cite{Story2018}. %\lvnote{uniformato rispetto a come citiamo gli altri paper} 
%the paper of Story et al. \cite{Story2018}. 
It is worth noticing that a privacy policy page could combine both coarse-grained (e.g., ``\texttt{contact\_information}") and fine-grained (e.g. ``\texttt{contact\_email\_address}") PSI.

\begin{table}[!h]
\scriptsize
\centering
\begin{tabularx}{0.5\textwidth}{>{\arraybackslash}m{2.6cm}>{\arraybackslash}m{0.4cm}>{\arraybackslash}m{2cm}>{\centering\arraybackslash}m{0.5cm}>{\centering\arraybackslash}m{0.5cm}>{\centering\arraybackslash}m{0.5cm}}
\toprule
    % P = PRECISION
    % A = ACCURACY
    % R = RECALL
    \textbf{CR Checker task} & \textbf{Model} & \textbf{Best Parameters}& \textbf{P} & \textbf{A} & \textbf{R}   \\
    \midrule
        \rowcolor{Gray}
        \textbf{contact\_address} & RF  & entropy, log2, 1000 & $85\%$ & $85\%$ & $85\%$\\
        
        \textbf{contact\_city} & RF  & entropy, log2, 100 & $73\%$ & $73\%$ & $73\%$\\
        
        \rowcolor{Gray}
        \textbf{contact\_email\_address} & SVM  & 1, 0.1, rbf & $77\%$ & $77\%$ & $77\%$\\
        
        \textbf{contact\_information} & SVM  & 0.1, linear & $82\%$ & $82\%$ & $82\%$\\
        
        \rowcolor{Gray}
        \textbf{contact\_password} & RF  & entropy, auto, 100 & $83\%$ & $83\%$ & $83\%$\\
        
        \textbf{contact\_phone\_number} & LR  & 5, 20, l2, 10, lbfgs & $78\%$ & $78\%$ & $78\%$\\
        
        \rowcolor{Gray}
        \textbf{contact\_postal} & MNB  & 0.5, True & $80\%$ & $80\%$ & $80\%$\\
        
        \textbf{contact\_zip} & AB  & SAMME, 500 & $81\%$ & $81\%$ & $81\%$\\
        
        \rowcolor{Gray}
        \textbf{demographic\_age} & RF  & gini, auto, 50 & $83\%$ & $83\%$ & $83\%$\\
        
        \textbf{demographic\_gender} & AB  & SAMME.R, 265 & $78\%$ & $78\%$ & $78\%$\\
        
        \rowcolor{Gray}
        \textbf{demographic\_information} & RF  & gini, auto, 150 & $82\%$ & $82\%$ & $82\%$\\
        
        \textbf{identifier\_ad\_id} & RF  & entropy, auto, 10 & $83\%$ & $83\%$ & $83\%$\\
        
        \rowcolor{Gray}
        \textbf{identifier\_cookie} & SVM  & 1, 1, rbf & $88\%$ & $88\%$ & $88\%$\\
        %\textbf{identifier\_cookie\_or\_similar\_tech} & SVM  & $96.2\%$\\
        
        \textbf{identifier\_device} & RF  & gini, auto, 50 & $79\%$ & $79\%$ & $79\%$\\
        
        \rowcolor{Gray}
        \textbf{identifier\_imei} & RF  & entropy, log2, 50 & $91\%$ & $91\%$ & $91\%$\\
        
        \textbf{identifier\_imsi} & RF  & gini, auto, 5 & $100\%$ & $100\%$ & $100\%$\\
        
        \rowcolor{Gray}
        \textbf{identifier\_information} & RF  & gini, log2, 10 & $79\%$ & $79\%$ & $79\%$\\
        
        \textbf{identifier\_ip\_address} & RF  & entropy, log2, 1000 & $73\%$ & $73\%$ & $73\%$\\
        
        \rowcolor{Gray}
        \textbf{identifier\_mac} & RF  & gini, auto, 50 & $88\%$ & $88\%$ & $88\%$\\
        
        \textbf{identifier\_sim\_serial} & MNB  & 1.5, True & $100\%$ & $100\%$ & $100\%$\\
        
        \rowcolor{Gray}
        \textbf{identifier\_SSID\_BSSID} & RF  & gini, auto, 50 & $92\%$ & $92\%$ & $92\%$\\
        
        \textbf{location\_bluetooth} & RF  & entropy, log2, 10 & $83\%$ & $83\%$ & $83\%$\\
        
        \rowcolor{Gray}
        \textbf{location\_cell\_tower} & RF  & gini, auto, 100 & $87\%$ & $87\%$ & $87\%$\\
        
        \textbf{location\_gps} & RF  & gini, log2, 300 & $82\%$ & $82\%$ & $82\%$\\
        
        \rowcolor{Gray}
        \textbf{location\_information} & SVM & 0.01, linear & $73\%$ & $73\%$ & $73\%$\\
        
        \textbf{location\_ip\_address} & MNB  & 0.5, False & $78\%$ & $78\%$ & $78\%$\\
        
        \rowcolor{Gray}
        \textbf{location\_wifi} & RF  & entropy, auto, 25 & $82\%$ & $82\%$ & $82\%$\\
        
        \textbf{performed\_not\_performed} & RF  & gini, auto, 500 & $96\%$ & $96\%$ & $96\%$\\
        
        \rowcolor{Gray}
        \textbf{third\_party\_first\_party} & RF  & gini, log2, 300 & $98\%$ & $98\%$ & $98\%$\\
        
\bottomrule
\end{tabularx}
\caption{CR Checker tasks: model parameters and evaluation in terms of precision (P), accuracy (A), and recall (R).}%\dcnote{Spero vada meglio}
\label{tab:crevaluator_implementation_results}
\end{table}

%\paragraph*{\textbf{Dynamic Analysis}}
\paragraph*{\textbf{Dynamic Testing Environment}}
%\toolname~tests the Android apps using an Android device emulator, based on the Android x86 project. 
The app is executed on an emulated x86\footnote{\url{https://www.android-x86.org/}} device based on Android 6.0 with root permissions, and equipped with Frida, a dynamic code instrumentation toolkit\footnote{\url{https://frida.re/docs/home/}}. 
The DTE module is attached to the emulator and orchestrates the installation, execution, and stimulation of the app under test by leveraging the Android Debug Bridge (ADB)\footnote{https://developer.android.com/studio/command-line/adb}.
The \textit{Privacy API Monitor} relies on Frida to instrument the app at runtime.
This module is configured to log only the privacy-related API identified by the \textit{App Analyzer}, and to analyze an app for ten minutes at most.

\paragraph*{\textbf{Privacy Policy Page (3P) Detector}}
The 3P Detector stimulates the app using the DTE module and extracts the XML of the app screens using the UI Automator tool\footnote{https://developer.android.com/training/testing/ui-automator}. 
Then, the module pre-processes the policy text using the NLTK\footnote{https://www.nltk.org/} libraries, and relies on a multilayer perceptron (MLP) based text classifier based on the TensorFlow library \cite{tensorflow} to identify the policy page.
The MLP is composed of two hidden layers made by 64 perceptrons each. The activation function of the output layer is a sigmoid function, and the loss function used to train the model is a binary cross-entropy function. Moreover, we set the learning rate to $0.0001$,  the number of epochs to $1000$, a batch size to $128$, and a dropout rate to $0.2$ in order to prevent overfitting. %\arnote{Sono valori abbastanza standard nel settore, ho aggiunto che il dropout serve per evitare l'overfitting} \dcnote{Andrea magari qua dai tu una spiegazione dei parametri}.\amnote{O almeno intuizione del perche' sono scelte ragionevoli/cool/standard.}  

We leveraged the APP-350\cite{Zimmeck2019} and a subset of the News Summary\footnote{\url{https://github.com/sunnysai12345/News_Summary}} datasets - labeled as ``\texttt{policy}" and ``\texttt{not\_policy}" respectively - as a basis to build a new dataset for training and validating the MLP model.  
%To have a usable dataset by our MLP model, 
%As output, we obtained a database suitable for the MLP model.
Then, we carried out the following operations on this dataset: i) we tokenized the documents in n-grams (with $n=1$ and $n=2$), ii) we computed the importance of each n-gram using the $tf{-}idf$ function \cite{ramos2003tfidf} and iii) we selected the best $20k$ n-grams based on the ANOVA F-value \cite{girden1992anova} statistical test.
%\dcnote{Aggiungere citazione e dire che è un test statistico}

The output of the MLP model is a \textsl{probability score} that indicates the likelihood that the analyzed privacy page belongs to the ``\texttt{policy}" class.
We defined a confidence threshold to 90\%, i.e., the page is classified as ``\texttt{not\_policy}" if the score is less than the threshold; otherwise, it is classified as  ``\texttt{policy}".
We selected a high threshold in order to automatically discard all text pages which may resemble a privacy policy page (e.g., the ``Terms \& Conditions'' pages), thereby reducing the likelihood of false positives.

\paragraph*{\textbf{Technical Requirement (TR) Tester}}
%As with the P3 Detector module, the TR Tester is implemented as Python module.
The TR Tester interacts with the emulator using the DTE module and the UI Automator tool to evaluate the compliance of the privacy policy page w.r.t. the list of TRs (Table \ref{tab:google_play_guidlines}). 
As described in Section \ref{sec:methodology}, we defined a maximum number of $20$ actions to check the compliance with \texttt{TR2}, and we set a timeout of $10$ seconds for \texttt{TR6}.

\paragraph*{\textbf{Content Requirement (CR) Checker}}
The CR Checker analyzes each sentence in a privacy policy page with the aim to $i$) identify the set of PSI therein, $ii$) check whether the sentence is affirmative or not (e.g., ``We access your contacts" or ``We do not access your contacts"), and $iii$) verify if the PSI is accessed by third-party libraries.
%\end{enumerate}
%The CR Evaluator a common phase that preprocesses the data. 
The CR Checker pre-processes the privacy policy page by splitting it into sentences using the NLTK libraries.
Following the same approach of 3P Detector, the sentences are thus tokenized in n-grams (n=1 and n=2). Then, the $tf-idf$ function is applied to evaluate the importance of each n-gram.
%After the common phase, each task was developed independently of the others.
Moreover, the CR Checker leverages 27 ML models to identify the PSI in the sentence (i.e., one for each PSI).
Finally, the CR Checker leverages two ad hoc binary classifiers, i.e., \textsl{performed\_not\_performed} and \textsl{third\_party\_first\_party}, to identify whether i) a sentence is affirmative, and ii) the third-party libraries access PSI, respectively.

We selected and tested a set of ML algorithms (i.e., \textsl{MultinomialNB}, \textsl{RandomForest}, \textsl{SVM}, \textsl{kNN}, \textsl{LogisticRegression}, \textsl{DecisionTree}, and \textsl{AdaBoost}) in order to find the best model for each CR Checker task.
In addition, we carried out a hyperparameters optimization phase using a grid search strategy (i.e., exhaustive searching strategy) to find the best parameters for all ML algorithms; the tested parameters are summarized in Table \ref{tab:ml_algorithm_and_parameters}.
% In details, for MultinomialNB (MNB) algorithm we tuned the \textit{alpha} and \textit{fit\_prior} parameters, for RandomForest (RF) the \textit{critierion},  \textit{max\_features} and \textit{n\_estimators} parameters, for SVM the \textit{C}, \textit{gamma}, and \textit{kernel} parameters, for LogisticRegression (LR) the \textit{C}, \textit{max\_iter}, \textit{penalty}, and \textit{solver} parameters, for kNN \textit{n\_neighbors}, \textit{weights}, and \textit{algorithm} parameters, for DecisionTree (DT) the \textit{criterion} and \textit{splitter} parameters, and for AdaBoost (AB) the \textit{algorithm} and \textit{estimators} parameters. \amnote{Facciamo una piccola tabella? è più leggibile e da quello che capisco non dovrebbe overlapparsi con Tab. IV. Inoltre fornisce più hype alla parte ML, vista la venue.}\dcnote{tabella fatta, cancelliamo il testo?}

\begin{table}[h]
\footnotesize
\centering
\begin{tabularx}{0.5\textwidth}{>{\arraybackslash}m{4cm}>{\arraybackslash}m{4.9cm}}
\toprule

\textbf{ML Algorithm} & \textbf{Parameters}  \\
    \midrule
        \textbf{MultinomialNB (MNB)} & \textit{alpha}, \textit{fit\_prior} \\
       \textbf{RandomForest (RF)} & \textit{critierion},  \textit{max\_features}, \textit{n\_estimators}\\
        \textbf{Support Vector Machine (SVM)}  & \textit{C}, \textit{gamma}, \textit{kernel}  \\
        \textbf{Logistic Regression (LR)} & \textit{max\_iter}, \textit{penalty}, \textit{solver}  \\
        \textbf{k-Nearest Neighbors (kNN)} & \textit{n\_neighbors}, \textit{weights}, \textit{algorithm}  \\
       \textbf{Decision Tree (DT)} & \textit{criterion}, \textit{splitter}   \\
        \textbf{AdaBoost (AB)} & \textit{algorithm}, \textit{estimators}  \\
            
\bottomrule
\end{tabularx}
\caption{ML algorithms and their tested parameters.}
\label{tab:ml_algorithm_and_parameters}

\end{table}

The set of the best  parameters for each model, as well as the evaluation of the models in terms of precision (P), accuracy (A), and recall (R), are reported in Table \ref{tab:crevaluator_implementation_results}.

\section{Experimental Results}\label{sec:experimental_results}
We empirically assessed the reliability of the proposed methodology by systematically analyzing a dataset of $5,473$ apps with \toolname{}. 
Such apps are the top free Android apps ranked by the number of installations and average ratings according to Androidrank \cite{AndroidRank}, and have been downloaded from the Google Play Store between Dec. 2019 and Jan. 2020. It is worth noticing that $4,567$ apps (i.e., $\approx 84.4\%$) have a valid link to a privacy policy page on the Google Play Store.
Our experiments were conducted using a laptop equipped with an Intel Core i7-3770@3.40 GHz, 16GB RAM, and Ubuntu 18.04. 

\subsection{Overview of Apps}\label{sec:overview_results_app}

\toolname{} allowed categorizing the dataset according to the privacy-sensitive permissions and the usage of third-party libraries for accessing PSI.

The distribution of the privacy-relevant permissions in the app dataset is shown in Table \ref{tab:permissions_app_analyzed}.
The most requested privacy-sensitive permission is \texttt{ACCESS\_\-WIFI\_\-STATE} which allows accessing information about Wi-Fi networks and enables the extraction of tracking details about the users as explained in \cite{wifiState}. Other widely-used permissions include
\texttt{ACCESS\_\-FINE\_\-LOCATION} and \texttt{ACCESS\_\-COARSE\_\-LOCATION} that enable the access to GPS location and \texttt{READ\_\-PHONE\_\-STATE} that gives access to the phone state, including the phone number and information on the cellular network. 
A complete description of all Android permissions, including those listed in Table \ref{tab:permissions_app_analyzed}, can be found in \cite{AndroidManifest}.

%OLD version
%The distribution of the privacy-relevant permissions in the app dataset is shown in Table \ref{tab:permissions_app_analyzed}.
%The most requested privacy-sensitive permissions is the \texttt{ACCESS\_\-WIFI\_\-STATE} which allows to geolocalize the user indirectly. 
%\texttt{ACCESS\_\-FINE\_\-LOCATION} and  \texttt{ACCESS\_\-COARSE\_\-LOCATION} allow the apps to access precise or approximate location respectively.
%The \texttt{READ\_\-PHONE\_\-STATE} allows the apps to read the phone state, including the phone number and information on the cellular network. 
%\lvnote{qua rimuovere la parte in cui spiego cosa sono e fare una cosa più del tipo: come visibile nella tabella, i permessi related to PSI più usati includono l'accesso alle info wifi (ACCESS), alla memoria esterna e alle coordinate GPS}
%A complete description of all Android permissions, including those listed in 
%\arnote{the rimosso} 
%Table \ref{tab:permissions_app_analyzed}, can be found in \cite{AndroidManifest}.

\begin{table}[h]
\footnotesize
\centering
\begin{tabularx}{0.4\textwidth}{>{\arraybackslash}m{3.2cm}>{\centering\arraybackslash}m{1.5cm}>{\centering\arraybackslash}m{1.5cm}}
\toprule
\textbf{Permission} & \textbf{Percentage} & \textbf{Ratio}  \\
    \midrule
        \texttt{ACCESS\_WIFI\_STATE} & $47.1\%$ & \multicolumn{1}{r}{$2579/5473$} \\
        \texttt{READ\_EXTERNAL\_STORAGE} & $45.7\%$ & \multicolumn{1}{r}{$2503/5473$} \\
        \texttt{ACCESS\_FINE\_LOCATION} & $22.8\%$ & \multicolumn{1}{r}{$1250/5473$} \\
        \texttt{READ\_PHONE\_STATE} & $22.5\%$ & \multicolumn{1}{r}{$1233/5473$} \\
        \texttt{ACCESS\_COARSE\_LOCATION} & $21.4\%$ & \multicolumn{1}{r}{$1157/5473$} \\
        \texttt{CAMERA} & $18.8\%$ & \multicolumn{1}{r}{$1031/5473$}\\
        \texttt{GET\_ACCOUNTS} & $17.2\%$ & \multicolumn{1}{r}{$942/5473$} \\
        \texttt{RECORD\_AUDIO} & $11.1\%$ & \multicolumn{1}{r}{$606/5473$} \\
        \texttt{READ\_CONTACTS} & $9.5\%$ & \multicolumn{1}{r}{$518/5473$} \\
        \texttt{CALL\_PHONE} & $3.9\%$ & \multicolumn{1}{r}{$212/5473$} \\
        \texttt{READ\_CALENDAR} & $2.0\%$ & \multicolumn{1}{r}{$111/5473$} \\
        \texttt{READ\_SMS} & $0.8\%$ & \multicolumn{1}{r}{$43/5473$}\\
        \texttt{RECEIVE\_SMS} & $0.8\%$ & \multicolumn{1}{r}{$42/5473$} \\
        \texttt{READ\_CALL\_LOG} & $0.7\%$ & \multicolumn{1}{r}{$38/5473$} \\
\bottomrule
\end{tabularx}
\caption{Distribution of privacy-related permissions in the experimental dataset.}
\label{tab:permissions_app_analyzed}
\end{table}

The distribution of third-party libraries for analytics and advertising is shown in Table \ref{tab:ads_and_analytics_library}.
It is important to emphasize that a single PSI can be shared with one or more third-party libraries, as an app can import any number of third-party libraries.
It is also worth pointing out that the most widespread libraries belong to Google (i.e., Google Ads, Google Firebase Analytics, Google DoubleClick, Google CrashLytics, and Google Analytics) and Facebook (i.e., Facebook Ads and Facebook Analytics).

\begin{table}[h]
\footnotesize
\centering
\begin{tabularx}{0.385\textwidth}{>{\arraybackslash}m{3cm}>{\centering\arraybackslash}m{1.5cm}>{\centering\arraybackslash}m{1.5cm}}
\toprule

\textbf{Library Name} & \textbf{Percentage} & \textbf{Ratio}  \\
    \midrule
        Google Ads & $82.8\%$ & \multicolumn{1}{r}{$4534/5473$} \\
        Google Firebase Analytics & $63.1\%$ & \multicolumn{1}{r}{$3454/5473$} \\
        Google DoubleClick  & $55.6\%$ & \multicolumn{1}{r}{$3044/5473$} \\
        Google CrashLytics & $40.2\%$ & \multicolumn{1}{r}{$2201/5473$} \\
        Facebook Ads & $29.3\%$ & \multicolumn{1}{r}{$1605/5473$} \\
        Google Analytics & $28.8\%$ & \multicolumn{1}{r}{$1581/5473$} \\
        Facebook Analytics & $26.9\%$ & \multicolumn{1}{r}{$1472/5473$} \\
        Unity3d Ads & $21.1\%$ & \multicolumn{1}{r}{$1156/5473$} \\
        Moat & $18.9\%$ & \multicolumn{1}{r}{$1036/5473$} \\
        Flurry & $16.3\%$ & \multicolumn{1}{r}{$894/5473$} \\
        Inmobo & $15.9\%$ & \multicolumn{1}{r}{$869/5473$}\\
        AppLovin & $14.8\%$ & \multicolumn{1}{r}{$810/5473$} \\
        Twitter MoPup & $14.3\%$ & \multicolumn{1}{r}{$783/5473$} \\
        Vungle & $12.9\%$ & \multicolumn{1}{r}{$711/5473$} \\
        Integral Ad Science & $11.7\%$ & \multicolumn{1}{r}{$640/5473$} \\
        AdColony & $11.5\%$ & \multicolumn{1}{r}{$627/5473$} \\
            
\bottomrule
\end{tabularx}
\caption{Distribution of third-party libraries for advertising and analytics.}
\label{tab:ads_and_analytics_library}
\end{table}

\subsection{Success Rate and Performance Analysis}\label{sec:succesful_rate_and_performance_analysis}
\toolname{} was able to analyze $92.4\%$ of apps (i.e., $5057/5473$) successfully.
The analysis of the remaining $416$ apps failed due to one of the following reasons:

%a two-fold reason\lvnote{detta così in inglese significa che fallisce per entrambe le ragioni contemporaneamente}: 
\begin{itemize}
    
\item \textit{Technical}. The dynamic analysis of \toolname{} is based on an Android with root permissions, and mounted on an emulated x86 architecture equipped with the Houdini ARM translator library.
%\lvnote{troppi based in questa frase} 
However, some apps did not execute in an emulator or on a rooted device. Other apps failed due to compatibility issues related to the Houdini library.
%However, few apps failed the analysis for being not compatible with an emulated or rooted environment, while in other cases, the translator library raised compatibility exceptions.
% \item \textit{Hardware Architecture}. The emulator used by \toolname{} is based on the x86 architecture, however, some Android apps are executable only on devices based on ARM architecture. \amnote{Qualcuno potrebbe dire: si ma esistono le binary translation libraries. Perche' non le avete usate e messe nell'emulatore?}
\item \textit{Geographical}. Some apps (e.g., banking) executes only in specific locales.
\end{itemize}

Concerning performance, \toolname{} took $413$ hours for analyzing all the apps (i.e., $5473$), with a mean of $272$ seconds for each app, on a mid-level laptop, thereby suggesting that the approach is viable. Further optimizations like porting the \toolname{} approach on some Cloud IaS could be an interesting technical deployment to delve further that could consequently allow increasing the current thresholds for the dynamic analysis. %\lvnote{come mail il fog? mi sembra un pò una supercazzola}

%\amnote{Questa motivazione è weak. Sarebbe ok anche se fosse un ordine di grandezza maggiore. Se la nostra soluzione permette di verificare la compliance, immaginate Google o altri che problema possono avere a metterci il doppio  o a metterlo su un server. Vorrei la vostra opinione. Giriamo la cosa o smoothiamola. Consideriamo che abbiamo cmq upperboundato in maniera cheap le operazioni costose, come diciamo dopo.}
%This time includes all the steps necessary for \toolname{} to fully analyze the apps, including the time needed to start the emulator and install the app. 
%It is worth noticing that each analysis has been upper-bounded through two types of timeout. 
%The first one (see Section \ref{sec:methodology}) is related to the maximum number of actions that \toolname{}  executes to try reaching the privacy policy page (i.e., $20$ actions). 
%The second one is the maximum duration of each analysis: $10$ minutes. 
\begin{figure*}[!h]
    \centering
    \includegraphics[width=0.9\textwidth]{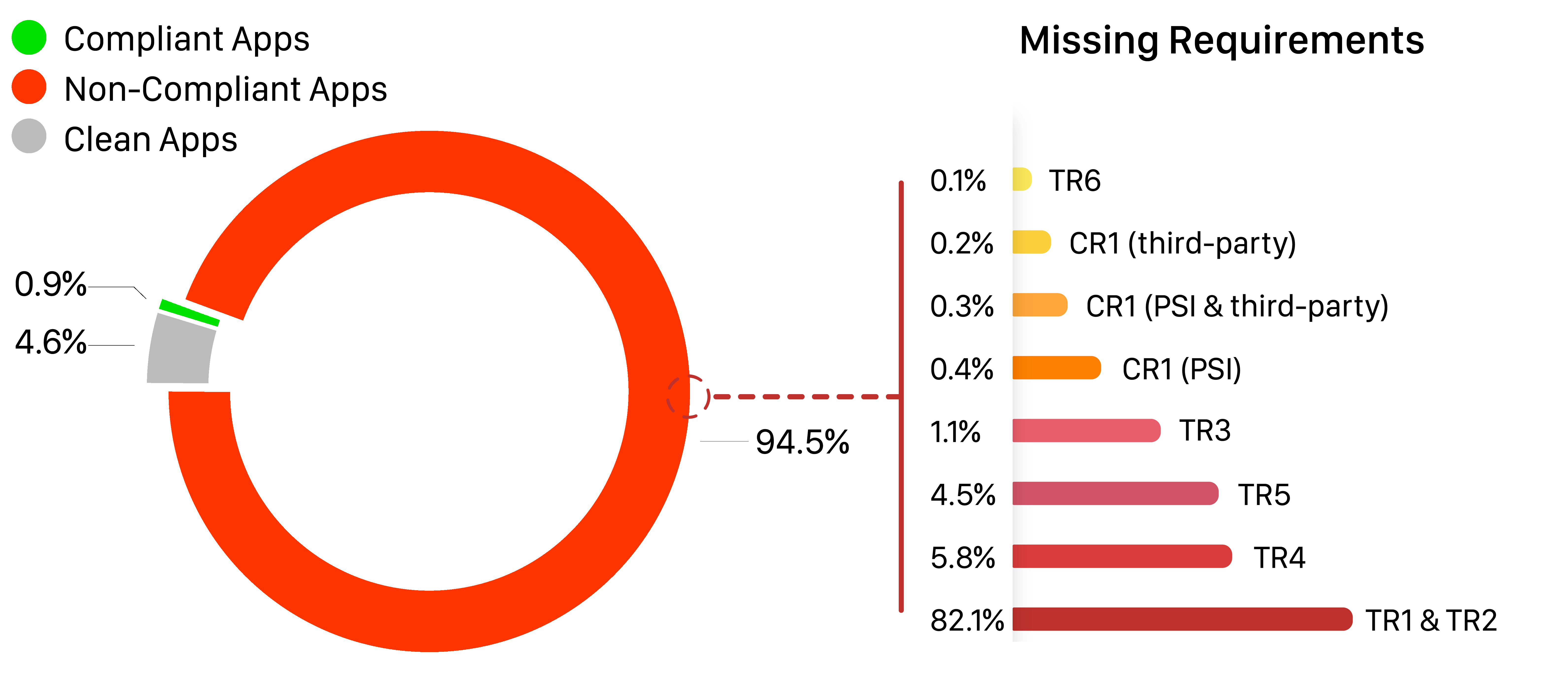}
    \caption{Overview of \toolname{} Results on 5057 apps.}
    \label{fig:3pdroid_results_overview}
\end{figure*}

\subsection{Analysis Results}
\label{sec:3pdroid_results}
The results obtained by \toolname{} are depicted in Fig. \ref{fig:3pdroid_results_overview} and indicate that only $5.5\%$ of the analyzed apps (i.e., $279/5057$) are compliant with the Google Play privacy guidelines. 

Among these, $4.6\%$ (i.e., $233/5057$) are \emph{clean} apps, i.e., apps that do not access any PSI, and thus do not require a privacy policy. 

It is worth pointing out how such a value suggests that only a minimal set of the current apps do not access any personal information. %, thereby highlighting the relevance of such kind of assessment.

The remaining $0.9\%$ (i.e., $46/5057$) are apps that actually provides a privacy policy page that fulfills the Google Play privacy guidelines.

The amount of non-compliant apps is worrisome ($94.5\%$, $4778/5057$), and suggests that the privacy problem could be more severe than foreseen in previous work. Furthermore, it is worth recalling that the vast majority of such apps have a valid link to a privacy policy page on the Google Play Store.
%\lvnote{qua non dobbiamo dare una valutazione personale ma descrivere i dati piuttsto dire che questa è una larga fetta} 

A more detailed analysis of this set shows that most of the apps ($82.1\%$, $4150/5057$) lack an internal privacy policy page - although they get access to PSI - and thus, they do not fulfill the \emph{TR1} and \emph{TR2} requirements. The remaining apps ($12.4\%$, $628/5057$) do not fulfill either \emph{TR3-TR6} or \emph{CR1}. It is worth pointing out that the analysis process is sequential and the TRs are evaluated before $CR1$. As a consequence, all apps that are tested against $CR1$ have been previously recognized as $TR3$-$TR6$ compliant, also.
%During the evaluation of those requirements, it is worth noticing that \toolname{} determines the not compliance of an app once it detects the first violation. To this aim, most of such apps do not pass through the entire assessment process.

Concerning TRs, $292$ apps collect PSI before an explicit acceptance of the privacy policy (\emph{TR4}), while $228$ assume that the user implicitly accepts the policy by leaving the privacy policy page (\emph{TR5}). Moreover, $54$ apps do not require an explicit acceptance from the user (\emph{TR3}), while $6$ of them have a self-expiring privacy policy page (\emph{TR6}).
Concerning the CRs, $22$ apps provide just a partial description of the PSI they collect during normal execution (i.e., \emph{CR1 PSI}), while $10$ apps do not warn the user about the usage of PSI-related third-party libraries (i.e., \emph{CR1 third-party}). 
Finally, $16$ apps do not fulfill any of the previous requirements (i.e., \emph{CR1 PSI \& third-party}).

\subsection{Manual Validation of \toolname{}}
We manually analyzed some subsets of apps, in order to assess the reliability of the machine learning-based components (i.e., 3P Detector and CR Checker) of \toolname{}.

Regarding 3P Detector, we manually analyzed a set of $1348/5057$ apps (i.e., $\approx 26.6\%$) made by two subsets. The first subset contains all $674$ apps in which 3P Detector recognized an internal privacy policy page. The latter is made by other $624$ apps randomly selected among the remaining ones (i.e., without a detected privacy policy page). 
%We leveraged such dataset to assess the accuracy of the 3P Detector module.
%The dataset is divided in two subset, made by all $674$ apps in which \toolname{} recognized a privacy policy page, and a set these apps plus an equal number of apps randomly selected among the remaining ones (i.e., without a privacy policy page). 
%We refer to  $A_p$ an app recognized by \toolname{} as having a privacy policy page, and  an app without a privacy policy page, $D$ is made by 674 $A_p$ and 674 $A_{\bar{P}}$, built as follows: 
%Such set is made by all $674$ apps recognized as having a privacy policy page (hereafter, $App_p$) plus an equal number of app randomly selected among those without a privacy policy page (hereafter, ).    
%We carried out a manual analysis of  results obtained by \toolname{} in order to check the reliability... the correctness of the modules based on machine learning.
%In detail, we manually checked the $674$ policies identified by the ML model, plus the pages of other $674$ randomly selected apps without PPP (Table \ref{tab:manuale_evaluation_of_3pdroid}) to prove the correctness of the 3P Detector module.
Regarding the first subset, the manual analysis showed that $648$ apps are actually policy pages (true positive - TP), while the remaining $26$ hosted a different kind of page (false positive - FP). Regarding the second subset, we found that $656$ apps actually lack a privacy-related content (true negative - TN), while the remaining $18$ apps have an undetected privacy policy page (false negatives - FN). 
%Globally, the evaluation carried out by 3P Detector failed $44$ out of $1348$, thereby reaching an accuracy of $96.7\%$.  

Concerning CR Checker, we took into consideration the $94$ apps, which fulfilled all TRs.
We manually analyzed all sentences (i.e., $2961$) contained in their privacy policy pages. 
CR Checker recognized $52/94$ apps as \textsl{CR1 (PSI)} compliant, and $64/94$ as \textsl{CR1 (third-party)} compliant. 
Regarding \textsl{CR1 (PSI)}, our manual analysis revealed that: i) $48/52$ are actually compliant (TP), while $4/52$ are not (FP). Furthermore, $28/42$ are indeed not compliant with \textsl{CR1 (PSI)} (TN), while $14/42$ revealed to be compliant (FN).  
%CR1 (PSI) compliant by \toolname{} (i.e., $42$),
Regarding \textsl{CR1 (third-party)}, we found that $63/64$ are TPs (and $1/64$ is a FP), while $24/30$ are TNs (and $6/30$ are FNs).
The results of the previous analysis indicate that the adopted classifiers have a good level of performance in terms of accuracy, sensitivity, specificity, and precision, as summarized in Table \ref{tab:manuale_evaluation_of_3pdroid}.

\begin{table}[h]
\scriptsize
\centering
\begin{tabularx}{0.5\textwidth}{>{\arraybackslash}m{0.8cm}>{\centering\arraybackslash}m{2cm}>{\arraybackslash}m{1.8cm}>{\centering\arraybackslash}m{1.2cm}>{\centering\arraybackslash}m{1.2cm}}
\toprule

 \multicolumn{2}{c}{\textbf{Metric}}  & \textbf{Model} & \textbf{Percentage} & \multicolumn{1}{r}{\textbf{Ratio}}  \\
    \midrule

    \multirow{3}{*}{\textbf{Accuracy}} & \multirow{3}{*}{ \scriptsize$\mathbf{\frac{TP + TN}{ TP + TN + FP + FN}}$}& 3P Detector & $96.7\%$ & \multicolumn{1}{r}{$1304/1348$}\\
    &&CR1 (PSI) & $80.9\%$ &  \multicolumn{1}{r}{$76/94$}\\
     &&CR1 (third-party) & $92.6\%$ &  \multicolumn{1}{r}{$87/96$}\\
    \midrule
    
     \multirow{3}{*}{\textbf{Sensitivity}} &  \multirow{3}{*}{ \scriptsize$\mathbf{\frac{TP}{ TP + FN}}$}& 3P Detector & $97.3\%$ &  \multicolumn{1}{r}{$648/666$}\\
    &&CR1 (PSI) & $77.4\%$ &  \multicolumn{1}{r}{$48/62$}\\
     &&CR1 (third-party) & $91.3\%$ &  \multicolumn{1}{r}{$63/69$}\\
    \midrule
    
     \multirow{3}{*}{\textbf{Specificity}} & \multirow{3}{*}{ \scriptsize$\mathbf{\frac{TN}{ TN + FP}}$}& 3P Detector & $96.2\%$ &  \multicolumn{1}{r}{$656/682$}\\
    &&CR1 (PSI) & $87.5\%$ &  \multicolumn{1}{r}{$28/32$}\\
     &&CR1 (third-party) & $96.0\%$ &  \multicolumn{1}{r}{$24/25$}\\
    \midrule
    
     \multirow{3}{*}{\textbf{Precision}} & \multirow{3}{*}{ \scriptsize$\mathbf{\frac{TP}{TP+ FP}}$}& 3P Detector & $96.1\%$ &  \multicolumn{1}{r}{$648/674$}\\
    &&CR1 (PSI) &  $92.3\%$ &  \multicolumn{1}{r}{$48/52$} \\
     &&CR1 (third-party) &  $98.4\%$ &  \multicolumn{1}{r}{$63/64$}\\

\bottomrule
\end{tabularx}
\caption{Performance of ML-based analysis in \toolname{}.}
\label{tab:manuale_evaluation_of_3pdroid}
\end{table}

\section{Conclusion}
\label{sec:conclusion}
In this paper, we introduced the first methodology which allows assessing the compliance of Android apps with the recently released Google Play privacy guidelines \textsl{at runtime}. Our approach can be combined with the previous proposals based on static analysis, in order to build more reliable analysis workflows for evaluating the access to PSI by Android apps. 
Our results suggest that the vast majority of actual Android apps (i.e., $95.4\%$ of the analyzed apps) access PSI, but just a negligible part of them (i.e., $\approx 1\%$) fully complies with the Google Play privacy guidelines. %\amnote{Mettere una frase per dire "c'è molto da fare". Ma potrebbe essere superfluo. Idee?}
%In this work, we have proposed a methodology for systematically evaluating the compliance of Android apps with the Google Play privacy guidelines. 
%The methodology combines static, dynamic analysis and machine learning techniques to evaluates both the technical and content requirements that are asked to apps that collect, use and share personal and sensitive information (PSI) of the user and the device. 
%The methodology enables us to extract all the app permissions and the third-party libraries that handle PSI and detect the presence of the required in-app disclosure, i.e., the privacy policy page.
%Then, we are able to evaluate its compliance with the technical requisites, e.g., the presence of an explicit acceptance item, and detect whether the app starts the collection of PSI prior to the consent of the user.
%Besides, our methodology exploits state-of-the-art machine learning techniques to recognize the app screen containing the policy page and evaluate the content of the privacy policy, i.e., the presence in the policy of all pieces of PSI used by the app.
%Another contribution of this work is the implementation of a PoC based on our methodology, which enables the automatic compliance evaluation of Android apps in a black-box fashion. To evaluate the reliability and the performance of the tool, we are analyzing $10,000$ apps downloaded from the Google Play Store in December 2019 and January 2020.

Future extensions of this work could be i) an extensive empirical assessment of the methodology in the wild, by analyzing a higher number of Android apps, ii) the evaluation of other machine learning techniques for the detection of the privacy policy page and the classification of its contents, in order to further improve the precision, the recall, and the accuracy of the analysis, iii) extend the number of supported third-party libraries.  %\amnote{Forse dire meglio ultimo punto.}
%\lvnote{si dice precision and recall @andrea?}
%\dcnote{volendo anche accuracy (sono le 3 misure)}

%Citation example \cite{IEEEhowto:IEEEtranpage}.

\bibliographystyle{IEEEtran}
\bibliography{bibliography}

\end{document}